\documentclass[reprint, superscriptaddress, preprintnumbers, nofootinbib, amsmath, amssymb, aps, prl]{revtex4-1}

\usepackage{amsmath}
\usepackage{amssymb}

\usepackage{amsthm}
\usepackage{nicefrac}
\usepackage{amsfonts}

\usepackage[markup=nocolor]{changes}
\usepackage{natbib}

\usepackage{graphicx}

\usepackage[normalem]{ulem}

\usepackage{dcolumn}
\usepackage{bm}

\usepackage{color}
\usepackage[usenames,dvipsnames,svgnames,table]{xcolor}

\usepackage{slashed}
\usepackage{empheq}
\newcommand{\bea}{\begin{eqnarray}}
\newcommand{\eea}{\end{eqnarray}}
\newcommand{\be}{\begin{equation}}
\newcommand{\ee}{\end{equation}}

\allowdisplaybreaks

\newcommand{\GNewt}{Newton coupling}
\newcommand{\GNewts}{Newton couplings}

\usepackage{hyperref}
\hypersetup{
colorlinks=true,
citecolor=blue,
citebordercolor=red,
linktoc=all,
linkcolor=blue,
urlcolor=blue
}

\def\0#1#2{\frac{#1}{#2}}

\def\paragraphdistance{-0.98ex}

\begin{document}

\title{How perturbative is  quantum gravity?}
 
 \author{Astrid Eichhorn}
    \email{a.eichhorn@thphys.uni-heidelberg.de}
\affiliation{Institut f\"ur Theoretische
  Physik, Universit\"at Heidelberg, Philosophenweg 16, 69120
  Heidelberg, Germany}
  
      \author{Stefan Lippoldt}
    \email{s.lippoldt@thphys.uni-heidelberg.de}
\affiliation{Institut f\"ur Theoretische
  Physik, Universit\"at Heidelberg, Philosophenweg 16, 69120
  Heidelberg, Germany}
  
    \author{Jan M.~Pawlowski}
    \email{j.pawlowski@thphys.uni-heidelberg.de}
\affiliation{Institut f\"ur Theoretische
  Physik, Universit\"at Heidelberg, Philosophenweg 16, 69120
  Heidelberg, Germany}
  
    \author{Manuel Reichert}
    \email{reichert@cp3.sdu.dk}
\affiliation{Institut f\"ur Theoretische
  Physik, Universit\"at Heidelberg, Philosophenweg 16, 69120
  Heidelberg, Germany}
\affiliation{CP$^3$-Origins, University of Southern Denmark, Campusvej 55, 5230 Odense M, Denmark}
  
   \author{Marc Schiffer}
    \email{m.schiffer@thphys.uni-heidelberg.de}
\affiliation{Institut f\"ur Theoretische
  Physik, Universit\"at Heidelberg, Philosophenweg 16, 69120
  Heidelberg, Germany}

\begin{abstract}
We explore asymptotic safety of gravity-matter systems, discovering indications for a near-perturbative nature of these
systems in the ultraviolet. Our results are based on the dynamical emergence of effective universality at the asymptotically
safe fixed point. Our findings support the conjecture that an asymptotically safe completion of the Standard Model with
gravity could be realized in a near-perturbative setting.
\end{abstract}

\maketitle

{\it Introduction---} How nonperturbative is quantum gravity? What
appears to be a technical question at the first glance, could actually
be critical for the conceptual understanding of quantum spacetime. The
nature of quantum gravity in the very early universe, its impact on
matter, and the prospects of potential observational tests depend on
the extent to which quantum spacetime is nonperturbative.

In a quantum field theoretic setting, the ultraviolet (UV) completion
of particle physics including gravity may be asymptotically safe
\cite{Weinberg:1980gg}, being governed by the interacting Reuter fixed
point in quantum gravity \cite{Reuter:1996cp}. In the past two decades
substantial nontrivial evidence has been collected for the existence
of this fixed point in gravity-matter systems. 
In parallel, the search for a universal emergence of spacetime encoded in
an interacting fixed point is also ongoing in other quantum gravity approaches.
Despite the impressive
plethora of results some pressing questions concerning the physical
nature of the fixed point have not been answered yet.

In \cite{Eichhorn:2018akn} the concept of {\it effective universality}
of gravitational couplings has been put forward. Within a
scalar-gravity system, the gravitational self-coupling and the
scalar-gravity coupling are in semi-quantitative agreement. These are
{\it avatars} of the Newton coupling that agree on the classical
level. Their near-agreement in the quantum theory suggests a
near-perturbative realization of diffeomorphism invariance in the
asymptotically safe UV regime. Simply put, if the fixed point lies in
a near-classical regime with small quantum fluctuations, these
avatars are nearly equal.  This result complements indications for a
near-perturbative fixed-point structure based on
near-canonical scaling dimensions \cite{Falls:2013bv}.  This highly
intriguing scenario would allow for the application of perturbative
methods in the transplanckian regime, which indeed provide indications
for the fixed point \cite{Niedermaier:2009zz}.

In this work we find indications for this scenario in
gravity-matter systems, including all types of Standard-Model
fields. We discover that effective universality
\cite{Eichhorn:2018akn} holds for the avatars of the Newton coupling
including the gravitational self-couplings and all minimal
gravity-matter couplings, see
\cite{Christiansen:2015rva,Meibohm:2015twa,Dona:2015tnf,Eichhorn:2017sok,Christiansen:2017cxa,
  Eichhorn:2018akn,Eichhorn:2018nda}. This suggests that the UV fixed
point lies in a near-perturbative regime. \\[\paragraphdistance]

{\it Effective universality---} In the absence of a cosmological constant, General Relativity with matter is
parameterized by one single coupling, the Newton coupling
$G_\text{N}$. It governs
both, the gravitational self-interaction as well as the gravity-matter
interactions. In the presence of quantum fluctuations the single
classical Newton coupling is promoted to potentially different running
couplings, related to the interaction vertices of gravitons with each
other, with their Faddeev-Popov ghosts and with matter. The
scale-dependence of these avatars of the Newton coupling is encoded in
their $\beta$-functions. This structure is familiar from gauge
theories such as QED or QCD.  For marginal, i.e., dimensionless
couplings the corresponding avatars exhibit two-loop universality due
to gauge symmetry. Hence in the perturbative regime all vertices can
be described by a single gauge coupling. The Newton coupling is not
marginal. Therefore universality of the distinct $\beta$-functions is
not automatic. Nevertheless, the underlying symmetry manifests itself
in relations between the various \GNewts, the Slavnov-Taylor
identities (STIs).  These can lead to significant differences between
the various avatars in a nonperturbative regime of the theory. An
example is given by Landau gauge Yang-Mills theory in the
infrared. There the three-gluon coupling even becomes negative while
the other couplings remain positive, see e.g.\
\cite{Cyrol:2016tym,Pelaez:2013cpa,Aguilar:2013vaa,Blum:2014gna,Eichmann:2014xya}.

\begin{figure*}[!t]
\includegraphics[width=\linewidth]{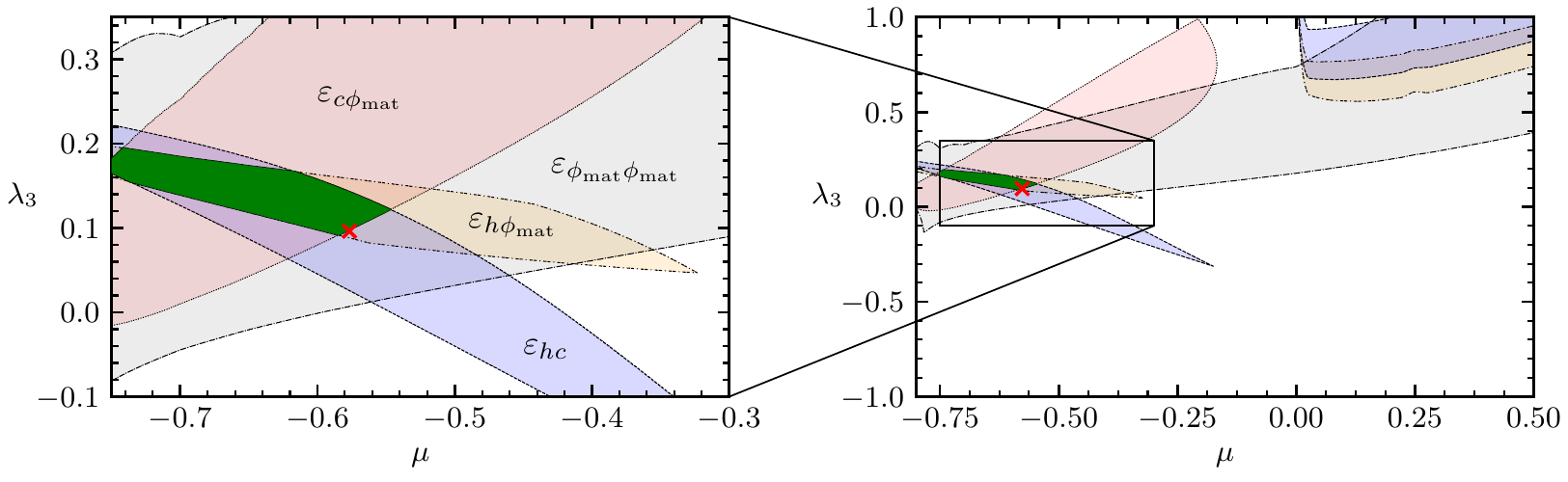}
\vspace{-0.6cm}
\caption{Regions in the $\mu$-$\lambda_{3}$--plane at $G=0.58$ where
  different sectors of the gravity-matter system are effectively
  universal ($\phi_{\text{mat}}= (\varphi,\psi,A)$).  Green region:
  effective universality in all avatars of the Newton coupling.  Red
  cross: UV fixed point.}
\label{fig:effuni}
\end{figure*}

It is a key result of this work that the $\beta$-functions of the different
avatars of the Newton coupling agree semi-quantitatively in the
asymptotically safe UV regime in gravity-matter systems. This
result is summarized in \autoref{fig:effuni}, showing that a
region of effective universality exists in the space of couplings. 
The location of the fixed point falls into 
this region.  We interpret the non-trivial emergence of effective universality as a manifestation of the
near-perturbative nature of asymptotically safe gravity. 
This supports a rather appealing scenario: the residual interactions
in the UV are just strong enough to induce asymptotic safety, while
allowing for near-canonical scaling of higher-order operators. The
near-canonical scaling is indeed observed, see e.g.\
\cite{Falls:2013bv,Falls:2014tra}  as is the existence of the
fixed point in perturbative studies \cite{Niedermaier:2009zz}. 
Additionally, the asymptotically safe Standard Model
\cite{Shaposhnikov:2009pv,Harst:2011zx,Eichhorn:2017ylw,
Eichhorn:2017lry,Eichhorn:2017als,Eichhorn:2018whv}
favors a perturbative nature of the fixed point. \\[-1.85ex]

{\it Gravity-matter systems---}We start from the gauge-fixed
Einstein-Hilbert action with minimally coupled scalars and fermions
as well as gauge-fixed  gauge theory with
$N_{\rm v}$ gauge fields.  The classical Euclidean action reads
\begin{align} \notag
S  &=  \01{16 \pi G_{\rm N}} \int\! \mathrm d^4 x\, 
\sqrt{g}\,\left(2 \Lambda \! - \! R \right)+ S_\text{gf+gh,gravity} 
\\
& + \frac{1}{2} \, \sum_{i=1}^{N_{\rm s}} \int\! \mathrm  d^4 x\, 
\sqrt{g}\,g^{\mu \nu} \partial_{\mu}\varphi^i\partial_{\nu}\varphi^i
+ \sum_{j=1}^{N_{\rm f}} \int\! \mathrm d^4x \sqrt{g} \,\bar \psi^j \slashed \nabla\psi^j
\notag \\
& +  \frac{1}{2} \int\! \mathrm{d}^4 x 
\sqrt{g} \, g^{\mu \nu} g^{\rho\sigma} \, {\rm tr}
F_{\mu \rho} F_{\nu \sigma}
+ S_\text{gf+gh,gauge} \,, 
\label{eq:action}\end{align}
where $F_{\mu\nu}$ is the field-strength tensor of the gauge field
$A_\mu$.  The gravity gauge fixing is of the linear de-Donder type in
the Landau gauge limit.  We also use the Landau gauge in the
Yang-Mills sector.  For the covariant Dirac operator
$\slashed{\nabla}$ we use the spin-base invariant formulation
\cite{Weldon:2000fr,Gies:2013noa,Lippoldt:2015cea}.  Under the impact
of quantum gravity, Abelian and non-Abelian gauge theories can
approach a free fixed point
\cite{Daum:2009dn,Harst:2011zx,Folkerts:2011jz,Christiansen:2017gtg,Eichhorn:2017lry,
  Christiansen:2017cxa}, such that the corresponding gauge couplings
vanish, and the non-Abelian ghost sector decouples.  Hence for our
computation only the total number of gauge fields, $N_{\rm v}$, is
relevant.  We focus on $N_{\rm s}= N_{\rm v}= 2 N_{\rm f}=1$. 

Expanding the metric about a flat  background 
\begin{align} \label{eq:lin_split}
g_{\mu\nu} = \delta_{\mu\nu}+ \sqrt{ G_{\rm N}}\,h_{\mu\nu}\,,
\end{align}
schematically leads to interactions of the form
\begin{align} \label{eq:interactions}
 \notag
 \Gamma
 \sim{}& \! \int \! {\rm d}^{4} x \big(
 \sqrt{G_{{\rm N},h}} \, h (\partial h) (\partial h)
 + \sqrt{G_{{\rm N},c}} \, h (\partial \bar{c}) (\partial c)
 \\
 \notag
 {}& + \sqrt{G_{{\rm N},\varphi}} \, h (\partial \varphi) (\partial \varphi)
 + \sqrt{G_{{\rm N},\psi}} \, h \bar{\psi} \gamma \partial \psi
 \\
 {}& + \sqrt{G_{{\rm N},A}} \, h (\partial A) (\partial A)
 \big)+...\,,
\end{align}
where we replaced $G_{\rm N}$ by 
avatars of the \GNewt{} $G_{{\rm N},i}$
corresponding to the interactions,
$i \in \{ h, c, \varphi, \psi, A \}$. In addition to the Newton
couplings, the expansion of the cosmological constant term results in
a two-graviton coupling $\mu$ and a three-graviton coupling
$\lambda_3$ \cite{Christiansen:2014raa,Christiansen:2015rva}.\\[\paragraphdistance]

{\it $\beta$-functions---}We compute the $\beta$-functions with the
functional renormalization group (FRG), for general reviews see
\cite{Berges:2000ew, Pawlowski:2005xe, Gies:2006wv, Delamotte:2007pf,
  Rosten:2010vm, Braun:2011pp}. For gravity it was pioneered in the
seminal paper of \cite{Reuter:1996cp}, for reviews see 
\cite{Niedermaier:2006wt,Litim:2011cp,Reuter:2012id,
Percacci:2017fkn,Bonanno:2017pkg,Eichhorn:2017egq}.
Here we follow the setup in
\cite{Christiansen:2012rx,Christiansen:2014raa,Christiansen:2015rva,
  Meibohm:2015twa,Dona:2015tnf,Eichhorn:2017sok,Christiansen:2017cxa,
  Eichhorn:2018akn,Eichhorn:2018nda}.
The involved algebra is handled using the symbolic manipulation system 
{\small FORM} \cite{Vermaseren:2000nd,Kuipers:2012rf} and the FormTracer \cite{Cyrol:2016zqb} 
as well as the Mathematica package xAct 
\cite{Brizuela:2008ra,2008CoPhC.179..597M,2007CoPhC.177..640M,2008CoPhC.179..586M}.
We provide a Mathematica notebook containing the final $\beta$-functions in numerical form \cite{Beta_Funcs.nb}.

We work with dimensionless running couplings, e.g.,
$G_i=G_{{\rm N},i}\, k^2$ for all avatars of the
Newton coupling.  It is already instructive to examine a simplified form of
the $\beta$-functions for the different avatars of the \GNewt{},
\begin{align}\label{eq:simplebeta}
 \beta_{G_i}
 = 2\,G - a_{i} \, G^2  
  + \mathcal{O}(G^3)\,,
\end{align}
where all avatars of the \GNewt{} are identified, $G_{i} = G$.  The
coefficients $a_i$ of the quadratic terms read
 \begin{align}
 (a_{h}, \, a_{c}, \, a_{\varphi}, \, a_{\psi}, \, a_{A})
 = (3.7, \, 3.8, \, 2.9, \, 2.9, \, 2.6) \, , 
\end{align}
when evaluated at $\mu = -0.58$ and $\lambda_3=0.096$. These are
precisely the fixed-point values that we will present later, see
\eqref{eq:UVFP}.  Already in the present simple approximation, the
coefficients differ by no more than 32\%.  For quantitative studies, we use a
measure for the relative deviation of the $\beta$-functions introduced in 
\cite{Eichhorn:2018akn},
\begin{align}
\varepsilon_{ij}(G, \mu, \lambda_3)=
  \left|\frac{\Delta\beta_{G_i}-\Delta\beta_{G_j}}{\Delta\beta_{G_i}+
  \Delta\beta_{G_j}}\right|_{G_i=G_j=G}\,,
 \end{align}
where $i,j \in \{h,c,\varphi,\psi,A\}$. 
$\Delta \beta_{G_{i}}$ is the anomalous part of the $\beta$-function $\beta_{G_{i}}$
obtained by subtracting the canonical running,
\begin{align}
\Delta\beta_{G_i} = \beta_{G_i} - 2\,G_i\,.
\end{align}
In the case of effective universality $\varepsilon_{i j}$ is close to
zero.  Larger values of $\varepsilon_{i j}$ signal a stronger
deviation from effective universality. This measure can be applied
pairwise to the 10 distinct pairs of $\beta$-functions. Due to a
rather mild $G$ dependence around the fixed point with $G^*_{h} = 0.58$,
cf.~\eqref{eq:UVFP}, we focus the discussion on the
$\mu$-$\lambda_{3}$--plane at $G^*_{h}$.

\begin{figure}[!t]
\includegraphics[width=0.98\linewidth]{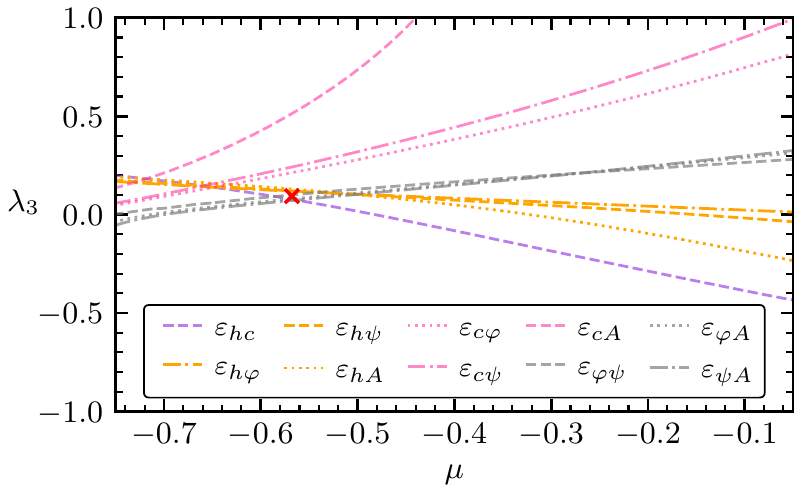}
\vspace{-0.3cm}
\caption{Lines with $\varepsilon_{ij}=0$
 in the $\mu$-$\lambda_{3}$--plane at $G = G_h^*$.
 The red star marks the UV fixed point, see \eqref{eq:UVFP}.
 }
\label{fig:epszero}
\end{figure}
A first nontrivial result concerns the existence of distinct lines where the
individual components of $\varepsilon_{ij}$ vanish in the $\mu$-$\lambda_{3}$--plane,
cf.~\autoref{fig:epszero}. These lines cross pairwise in a bounded region.
Moreover, the different crossing points lie near each other. This is
crucial and highly nontrivial, as the distinct $\varepsilon_{ij}=0$
lines have quite different slopes. Therefore, there is a priori no
reason to intersect pairwise in a relatively small region in the
$\mu$-$\lambda_{3}$--plane.  As we tentatively correlate the emergence
of effective universality, i.e., an agreement of the
$\beta$-functions, with a near-perturbative nature of the system, the
vicinity of the $\varepsilon_{ij}=0$ lines is a preferred region for the couplings.

In the gravity-matter system investigated in this work 
the interacting Reuter fixed point lies at 
\begin{align} \label{eq:UVFP}
 (G_h^* ,\, G_c^* ,\, &G_\varphi^* ,\, G_\psi^* ,\, G_A^* ,\, \mu^* ,\, \lambda_3^* )
 \\
 &= (0.58 ,\, 0.55 ,\, 0.74 ,\, 0.74 ,\, 0.84 ,\, -0.58 ,\, 0.096 ) \,. \notag 
\end{align}
The values of the
different vertex couplings are related by the STIs but they are not necessarily
identical. This potential difference is ignored in the ensuing
qualitative discussion where we assume full universality for the sake
of simplicity. 

The flow in the vicinity of the fixed point is governed by the
 critical exponents at linearized order of the $\beta$-functions.
The STIs entail that only two of our couplings are independent. 
They are related to the classical Newton coupling and the cosmological
constant. This renders only a subset of critical exponents physical.  In our
setting, the critical exponents turn out to be quantitatively similar,
and close to the values obtained by setting $G_i = G$ \emph{before}
calculating the stability matrix by taking derivatives of the $\beta$-function.
This constitutes another strong indication for effective
universality, and provides an estimate of ${\rm Re}\,\theta \approx 1.3 - 1.7$
for the physical critical exponent of the Newton coupling.

\begin{figure}[!t]
\includegraphics[width=0.92\linewidth]{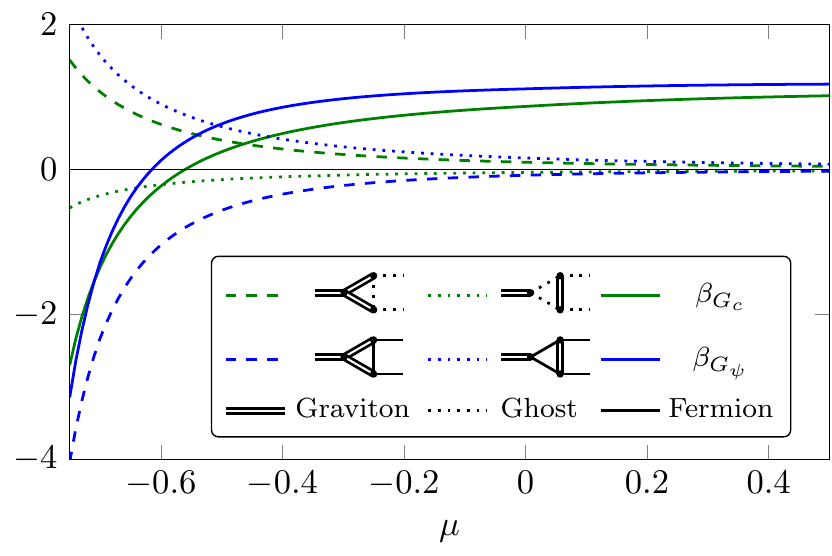}
\vspace{-0.3cm}
\caption{Nontrivial cancellations between different diagrams.  We
  show the $\mu$ dependence at $G = G_h^*$ and
  $\lambda_{3} = \lambda_3^*$ of $\beta_{G_{c}}$ and
  $\beta_{G_{\psi}}$, and of two contributing diagrams.
  }
\label{fig:StarFish}
\end{figure}
Crucially, our result requires nontrivial cancellations between the
diagrams that contribute to the distinct $\beta$-functions. It is
therefore not an automatic consequence of our choice of truncation
that is inspired by classical diffeomorphism invariance.  As a
specific example we highlight $\beta_{G_{c}}$ and
$\beta_{G_{\varphi}}$ in \autoref{fig:StarFish}.  There we find an
overall similarity of the $\beta$-functions while individual diagrams
do not agree. Such cancellations appear unlikely to be generated by
chance and we interpret them as a hint that the existence of the
$\varepsilon_{ij}=0$ lines and a fixed point in their vicinity is
indeed a nontrivial result.\\[-3.ex]

{\it Sources of deviations from effective universality---} For a
stringent assessment of the deviation of the fixed point from
$\varepsilon_{ij}=0$, we have to evaluate its possible origins. In the
present work we explore the Einstein-Hilbert truncation with minimally
coupled matter.  Yet, at an asymptotically safe fixed point,
higher-curvature couplings
\cite{Benedetti:2009rx,Falls:2013bv,Gies:2016con,Christiansen:2017bsy,deBrito:2018jxt}, nonminimal gravity-matter
couplings \cite{Eichhorn:2017sok,Eichhorn:2018nda} and
matter-self interactions \cite{Eichhorn:2011pc,Christiansen:2017gtg,Eichhorn:2017eht}
are also present.  Accordingly, our fixed-point values as well as the
location of the $\varepsilon_{ij}=0$ lines are subject to a systematic
error $\delta \varepsilon$. A comparison of the fixed point to the
$\varepsilon_{ij}=0$ lines is only meaningful within
$\delta \varepsilon$. A rough estimate for the systematic error --
strictly speaking an estimate for a lower bound on it -- can be
obtained by comparing changes in fixed-point values under extensions
of the truncation. Specifically, we compare a state-of-the-art study
\cite{Denz:2016qks} with a previous work in the same scheme
\cite{Meibohm:2015twa} to obtain differences in fixed-point values
$\delta G_h, \delta \mu, \delta \lambda_3$. The relative variation of
fixed-point values is similar in other extensions of truncations, see,
e.g., \cite{Eichhorn:2017sok,Knorr:2017fus}.  The average
$\delta \varepsilon$ of the $\delta \varepsilon_{ij}$ is given by
\begin{align}
  \delta \varepsilon \! = \!
  \frac1{10}  \sum\limits_{\substack{i, j \\ \! j \scalebox{0.7}{$ \, < \, $} i}} 
  \left[
  \Big|\frac{\partial \varepsilon_{ij}}{\partial G} \delta G_{h}\Big|
  \! + \! \Big|\frac{\partial \varepsilon_{ij}}{\partial \mu}
  \delta \mu\Big|
  \! + \! \Big|\frac{\partial \varepsilon_{ij}}{\partial \lambda_3}
  \delta \lambda_3\Big|
  \right]_{\!\!\! \substack{
  \scalebox{0.8}{$
 \begin{aligned}
   G ={}& G^* \\[-0.18cm] \mu ={}& \mu^* \\[-0.13cm]
   \lambda_{3} ={}& \lambda_3^*
 \end{aligned}
 $}
 }},
\end{align}
resulting in $\delta \varepsilon \approx 0.2$. 
As a key result, we stress that
$\varepsilon_{i j} \approx 0.2$ is compatible with effective
universality within this estimate of the error $\delta \varepsilon$,
cf.~contours in~\autoref{fig:epszero}. Accordingly, at the UV fixed point,
the avatars of the Newton coupling are compatible with effective universality.

The presence of higher-order operators results in a second source of
deviations of a more involved nature: It is rooted in the challenge of
projecting correlation functions onto specific operators and the
related couplings, e.g., the avatars of the Newton coupling.  For
instance, at the level of the graviton three-point function, the terms
contributing to our result for $\beta_{G_h}$ include $\sqrt{g}R$ and
$\sqrt{g}R_{\mu\nu}R^{\mu\nu}$, both expanded to third order in $h$.
In the present gauge-fixed and regularized setting, one faces the
additional challenge to account for nondiffeomorphic operators.

The higher-order contributions in the three-graviton and the
graviton-matter vertices are \emph{not} related to each other, as they
are linked to distinct classically diffeomorphism invariant operators
such as, e.g., $\sqrt{g}R_{\mu\nu}R^{\mu\nu}$ vs
$\sqrt{g}R^{\mu\nu}\partial_{\mu}\varphi \partial_{\nu}\varphi$. From
the observed small value of the $\varepsilon_{ij}$, we conclude that
such higher-order operators, which could spoil effective universality
completely, have a subleading impact.

Let us elucidate this point with an explicit example: Assume for the
moment, that our evaluation of $\beta_{G_h}$ would lead to
$2 G + \Delta \beta_{G_{h}} + \delta \beta_{\rm Ric}$. Here,
$\delta \beta_{\rm Ric}$ is an additional part of similar magnitude as
$\Delta \beta_{G_{h}}$, that originates from the running of the
$\sqrt{g} R_{\mu\nu}R^{\mu\nu}$ coupling and contributes to
$\beta_{G_h}$ due to our (non-diagonal) projection procedure. Even
assuming perfect agreement between all actual $\beta_{G_i}$, our
result for $\varepsilon_{h i}$ would be greater than $0.3$. The
observation that all $\varepsilon_{ij}$ satisfy
$\varepsilon_{ij}\lesssim0.2$ can tentatively be interpreted as a hint
for the subleading nature of higher-curvature couplings. Accordingly,
our projection prescription isolates the various \GNewts{} without a
large 'contamination' from higher-order terms. At the same time, this
suggests that the 'backreaction' of these specific higher-order terms,
once included in a truncation, should be small, as indeed observed in
several approximations, e.g., \cite{Falls:2013bv,Falls:2014tra}.\\[\paragraphdistance]

{\it Modified Slavnov-Taylor identities---}In a gauge-fixed setting,
the fluctuation couplings are related by nontrivial Slavnov-Taylor
identities (STIs).  In the flow-equation setup, the regulator function
is quadratic in the fluctuation fields and further breaks the
diffeomorphism invariance. This turns the STIs into modified STIs
(mSTIs) that now contain explicit regulator contributions, see e.g.\
\cite{Ellwanger:1995qf,Reuter:1996cp,Pawlowski:2005xe,
  Pawlowski:2003sk,Manrique:2009uh,Donkin:2012ud}.  The mSTIs imply
that couplings that derive from the same classical structure differ at
the quantum level. Thus, $\varepsilon_{ij}=0$ is not to be expected in
a quantum setting, even in the absence of the systematic effects
discussed above. Since the mSTIs arise as a consequence of quantum
effects, the perturbative limit with vanishing couplings features
trivial mSTIs. In the nonperturbative regime of gauge theories, the
mSTIs become nontrivial with QCD being an excellent example, see e.g.\
\cite{Cyrol:2016tym,Cyrol:2017ewj}. If we ascribed the full difference
in the fixed-point values of the different avatars of the \GNewt{} to
nontrivial mSTIs then $\varepsilon_{i j}\approx0.2$ would translate
into a factor of roughly 0.7 between the fixed-point values.  Taking
our cue from QCD, where different avatars of the gauge coupling even
feature distinct signs in the nonperturbative regime and thus much
larger relative differences, we tentatively conclude that our results
imply a near-perturbative nature of the asymptotically safe fixed
point.  More specifically, the analogue of mSTIs in QCD, see
\cite{Cyrol:2017ewj}, suggests a grouping of fixed-point values into
the pair $\{G_h, G_c\}$ and the triple
$\{G_{\varphi}, G_{\psi}, G_A\}$, with a nontrivial contribution from
the mSTI differentiating between the former and the latter. This
grouping is indeed apparent in the fixed-point values in
\eqref{eq:UVFP}. In contrast, the 'matter-like' behavior of
$\varepsilon_{c\phi_\text{mat}}$ and $\varepsilon_{hc}$ away from the
fixed point, cf.~\autoref{fig:effuni}, is a direct consequence of the
diagrammatic structure underlying the $\beta$-functions.\\[\paragraphdistance]

{\it Implications---}We observe that the fixed point yields
$\varepsilon_{i j} \approx 0.2$, which is compatible with zero within
our estimate for the systematic error.  This entails a compatibility
of the fixed point with effective universality within our present
setup. The result has several important implications.

Firstly, it strongly suggests that the zero of the $\beta$-functions
observed above is actually a true fixed point, in contrast to a
truncation artefact. For the latter, there is no reason why delicate
cancellations as observed above should occur.  
Their presence strongly hints at the impact of a
symmetry principle. We view the delicate cancellations that occur in
all pairs of $\beta$-functions as strong evidence for the physical nature
of the asymptotically safe Reuter fixed point.

Secondly, we contrast the observed fixed-point structure with that of
a system where $h_{\mu\nu}$ is a spin-2 field living on a fixed
background. Then $h_{\mu\nu}$ would \emph{not} be part of the dynamic
spacetime geometry.  If it was just another 'matter' field protected
by shift symmetry, it would feature derivative couplings like those
that we have examined here. Yet, the absence of a symmetry principle
relating the distinct $G$'s would make a semi-quantitative agreement
of the fixed-point values rather unlikely. The presence of this
dynamical symmetry linked to the underlying dynamical diffeomorphism
invariance of quantum gravity is corroborated further by the
previously observed momentum locality of specific propagator and vertex
flows, \cite{Christiansen:2015rva,Denz:2016qks}. This property 
entails that the leading momentum dependence of different diagrams
cancel non-trivially at large momenta. Our result could thus be
interpreted as highlighting the geometric origin of $h_{\mu\nu}$ with
its corresponding spacetime diffeomorphism symmetry and background
independence.
 
Thirdly, we contrast the relative deviation of fixed-point values of
different avatars of the Newton coupling with significantly larger
deviations in nonperturbative QFTs.  The relation between different
avatars of the gauge coupling is carried by mSTIs, which allow large
deviations of these classically equal avatars in a nonperturbative
regime governed by large quantum fluctuations. The significantly
smaller differences between the different \GNewts{} can accordingly be
interpreted as a consequence of a near-perturbative regime, where
mSTIs simplify and a more `classical' notion of diffeomorphism
symmetry is realized.\\[\paragraphdistance]

{\it Conclusions---}In summary we find strong indications that quantum
gravity, though perturbatively nonrenormalizable, can be described
within a quantum field theory in a near-perturbative regime.  
The corresponding `small parameter' would be related to the 
deviation from canonical scaling of higher-order couplings.
Further evidence is required to back our discovery of potential
near-perturbativity of asymptotically safe gravity. The `minimally
quantum' nature of spacetime in the sense of small quantum
fluctuations provides a strong backing of the robustness and
reliability of various approximations commonly used in the literature
and could constitute a key cornerstone in the understanding of quantum
gravity.

At the physical level, our finding signifies that huge quantum
fluctuations of spacetime appear to be subdominant in the
ultraviolet. Our result could have important implications for other
quantum-field theoretic approaches to quantum gravity in which the
search for a continuum limit is an ongoing quest. This includes
approaches with causal and Euclidean dynamical triangulations
\cite{Ambjorn:2012ij,Ambjorn:2014gsa,Laiho:2016nlp}, the tensor track
\cite{Gurau:2010ba,Gurau:2016cjo,Eichhorn:2017xhy}, and Loop Quantum
Gravity/ Spin foams
\cite{Dittrich:2014ala,Dittrich:2016tys,Bahr:2016hwc}. In these
approaches not much is known about the underlying universality classes
in four dimensions yet. Our findings suggest that there is a
universality class for the continuum limit which is near-perturbative
in nature, resulting in a near-canonical spectrum of scaling
exponents. This observation could provide guidance in the
construction of a suitable dynamics in these models.

A scenario with a near-perturbative
UV completion also provides an important input for studies of the
very early universe, as well as the construction of phenomenological
models geared towards observational tests of quantum gravity.
Importantly, our discovery fits well to the perturbative nature of the
Standard Model at the Planck scale and could provide a compelling
picture of
Planckian dynamics controlled by a near-perturbative fixed point.\\[\paragraphdistance]

{\it Acknowledgements---} This work is supported by an
Emmy-Noether-grant of the DFG under AE/1037-1, is supported by EMMI,
is part of and supported by the DFG Collaborative Research Centre "SFB
1225 (ISOQUANT)", and is supported by the Danish National Research Foundation
under grant DNRF:90. MR also acknowledges funding from the HGSFP and
the IMPRS-PTFS.  AE also acknowledges support through a visiting
fellowship at Perimeter Institute. Research at Perimeter Institute is
supported by the Government of Canada through the Department of
Innovation, Science and Economic Development and by the Province of
Ontario through the Ministry of Research and Innovation.

\bibliography{EffUni_v2}

\begin{thebibliography}{73}
\bibitem [{Bet()}]{Beta_Funcs.nb}%
  \BibitemOpen
  \href@noop {} {}\bibinfo {note} {We provide a Mathematica notebook containing
  the numeric beta-functions.}\BibitemShut {Stop}%
\end{thebibliography}

\end{document}